\pdfoutput=1
\documentclass[sigconf,screen,nonacm]{acmart}
\AtBeginDocument{%
  }

\setcopyright{none}

\usepackage{xcolor} 
\usepackage[table]{xcolor} 

\newif\ifcomments
\commentsfalse     

\ifcomments
  \newcommand{\hans}[1]{\textcolor{brown}{\textit{[Hans: #1]}}}

\else
  \newcommand{\hans}[1]{} 
\fi

\begin{document}


\title{An Extension-Based Accessibility Framework for Making Blockly Accessible to Blind and Low-Vision Users}
\thanks{This paper has been accepted at the 1st International Workshop on User Interface and Experience for Software Engineering (UISE 2026), co-located with ICSE 2026. The final published version is available at \url{https://doi.org/10.1145/3786169.3788398}.}

\author{Rubel Hassan Mollik}
\email{RubelHassanMollik@my.unt.edu}
\orcid{0009-0004-9714-1877}
\affiliation{%
  \institution{University of North Texas}
  \city{Denton}
  \state{Texas}
  \country{USA}
}

\author{Vamsi Krishna Kosuri}
\email{vamsikrishnakosuri@my.unt.edu}
\orcid{0009-0003-5517-2248}
\affiliation{%
  \institution{University of North Texas}
  \city{Denton}
  \state{Texas}
  \country{USA}
}

\author{Hans Djalali}
\email{hansdjalali@my.unt.edu}
\orcid{0009-0003-9158-1575}
\affiliation{%
  \institution{University of North Texas}
  \city{Denton}
  \state{Texas}
  \country{USA}
}

\author{Stephanie Ludi}
\email{stephanie.ludi@unt.edu}
\orcid{0000-0002-3281-9219}
\affiliation{%
  \institution{University of North Texas}
  \city{Denton}
  \state{Texas}
  \country{USA}
}

\author{Aboubakar Mountapmbeme}
\email{aboubakarmountapmbeme@my.unt.edu}
\orcid{0000-0002-3758-5183}
\affiliation{%
  \institution{University of North Texas}
  \city{Denton}
  \state{Texas}
  \country{USA}
}

\renewcommand{\shortauthors}{Mollik et al.}

\begin{abstract}
Block-based programming environments (BBPEs) such as Scratch and Code.org are now widely used in K-12 computer science classes, but they remain mostly inaccessible to blind or visually impaired (BVI) learners. A major problem is that prior accessibility solutions have relied on modifications to the Blockly library, making them difficult to apply in existing BBPEs and thereby limiting adoption. We present an Extension-based Accessibility Framework (EAF) to make BBPEs accessible for BVI students. The framework uses a modular architecture that enables seamless integration with existing Blockly-based BBPEs. We present an innovative three-dimensional (3D) hierarchical navigation model featuring stack labeling and block numbering, mode-based editing to prevent accidental modifications, and WAI-ARIA implementation to ensure compatibility with external screen readers. We evaluated our approach by integrating the EAF framework into two BBPEs (covering 177 test cases) and conducting semi-structured interviews with four participants using VoiceOver, JAWS, and NVDA. Participants reported clearer spatial orientation and easier mental model formation compared to default Blockly keyboard navigation. EAF shows that modular architecture can provide comprehensive accessibility while ensuring compatibility with existing BBPEs.
\end{abstract}



\maketitle

\section{Introduction}
The Computer Science for All (CSForAll)~\cite{santo2019equity} initiative has boosted participation in computer science (CS) and computational thinking (CT) education among K-12 students. Through this initiative, block-based programming has become widespread in K-12 CS courses. Block-based programming~\cite{weintrop2019block} replaces the complex syntax of text-based programming with visual, drag-and-drop blocks, making learning programming more accessible for novices. Students find visual programming more engaging and learn fundamental programming logic and computational thinking faster~\cite{weintrop2017comparing}. Platforms such as Scratch~\cite{resnick2009scratch}, Microsoft MakeCode~\cite{ball2019microsoft}, and Code.org~\cite{codeorg2024blockly} have emerged as block-based programming environments (BBPEs), which have become an integral part of K-12 CS curricula~\cite{kaleliouglu2015new}.

However, the BBPEs are not accessible to learners who are blind or visually impaired (BVI) due to their over-reliance on visual elements and animations~\cite{sharfuddeen2020visual}. Moreover, drag-and-drop operations like moving blocks involve mouse-based interaction and visual feedback, which is challenging for BVI learners~\cite{mountapmbeme2020investigating}. As a result, BVI learners face barriers to participating in CS courses, which leaves them behind their sighted peers~\cite{mountapmbeme2021teachers}. Since many popular BBPEs, such as Scratch, MakeCode, and Code.org, are built on the Blockly library, making it accessible can benefit all Blockly-based platforms. 

Google Blockly recently added a default keyboard navigation~\cite{blockly2024keyboard} that offers limited support for keyboard interaction. However, it remains in an experimental stage and lacks screen reader compatibility, making it inaccessible to BVI learners. While promising prototypes such as Accessible Blockly~\cite{mountapmbeme2022accessible} and GridWorld~\cite{tabassum2024accessible} have shown potential for keyboard-based alternative navigation for BVI learners, these efforts primarily focus on small-scale programs and specific navigation tasks, with limited code editing support. Furthermore, these approaches inherit architectural constraints by modifying the Blockly codebase, which impedes interoperability and prevents integration into existing Blockly-based environments. Several implementations also rely on internal speech synthesis ~\cite{das2021accessible,tabassum2024accessible}, which conflicts with screen readers, creating compatibility issues.

This work is motivated by the need to develop an accessible solution for BBPEs that supports structured navigation and editing of complex programs while maintaining architectural modularity. We propose a novel accessibility framework that can be seamlessly integrated into Blockly-based systems without modifying the core codebase. This approach provides a scalable, interoperable, and screen reader–compatible solution, improving accessibility of BBPEs. 

The main contributions of this paper are as follows:

    \textbf{•} We present an Extension-based Accessibility Framework (EAF) that integrates accessibility features into Blockly-based environments. Unlike previous methods that require forking the Blockly repository, EAF extends across platforms and Blockly versions while maintaining compatibility and structural stability.
    
    \textbf{•} We introduce a three-dimensional (3D) navigation and editing model with stack labels and sequential block numbers. The design establishes a predictable structural navigation and editing experience for BVI learners.
    
    \textbf{•} We validate \emph{EAF}'s viability through integration testing with Blockly-based platforms, WCAG 2.1 accessibility compliance assessment, and expert evaluation. Our results demonstrate that EAF enables accessible block-based programming while working with external screen readers (NVDA, JAWS, VoiceOver).



\section{Related Work}
\label{sec:related_work}
Extensive research has been conducted to improve the accessibility of BBPEs for BVI individuals. Previous studies explored approaches that require Blockly code modification ~\cite{mountapmbeme2022accessible,das2021accessible,tabassum2024accessible}, and extension-based architectures remain underexplored even though they have potential for modularity and maintainability~\cite{ludi2017design}.

Google initially offered a text-based Blockly separately for BVI users, replacing the visual drag-and-drop block UI with a text-only interface~\cite{koushik2016accessible}. Ludi et al.~\cite{ludi2017design} then proposed an alternative to the visual drag-and-drop user interface, adding keyboard input and screen reader compatibility. This effort suggested adding functionalities via files that could be applied to Blockly-based projects. Accessible Blockly~\cite{mountapmbeme2022accessible} was developed following this approach, augmenting the Blockly codebase, which demonstrated success in navigating block-based code using keyboard and screen reader support. Google later added an experimental keyboard navigation in Blockly~\cite{blockly2024keyboard}, which makes blocks navigable using a keyboard but has no support for audio feedback or screen readers. Das et al.~\cite{das2021accessible} an accessible BBPE by modifying Blockly with synthetic speech, which does not provide audio feedback for all interactions and has no support for program editing. Moumita et al.~\cite{tabassum2024accessible} took a similar approach to introduce a web application that implements Code.org's Maze Game puzzles with a minimal set of blocks.

Other researchers explored creating entirely new systems or using fundamentally different design principles to achieve accessibility~\cite{bulovic2024designing,milne2018blocks4all, stefik2024accessible}. Blocks4All~\cite{milne2018blocks4all} introduced a successful touchscreen-based BBPE specifically designed for children with visual impairments that requires specific hardware (iPad) and a Dash robot for tangible output. However, the application is not suitable for sighted learners and is incompatible with mainstream BBPEs used by K-12 learners. Similarly, OctoStudio~\cite{bulovic2024designing} was designed with tinkerability for the Android platform, while it remains in progress. In contrast, Quorum Blocks~\cite{stefik2024accessible} proposes a new accessible block language incorporating a hardware-accelerated graphics pipeline, but it is not purely block-based.

As shown in the Table~\ref{tab:corrected_comparison}, prior studies primarily employ either Blockly library modifications or new systems. These approaches face fundamental trade-offs between architectural integration and full accessibility. Blockly modification creates isolated silos and version dependencies, blocking integration with mainstream BBPEs, while new systems differ from K-12 BBPE practices. Moreover, current implementations lack structured navigation models and provide only partial support for keyboard-only interaction and screen reader feedback. These gaps underscore the need for a modular architecture with comprehensive accessibility support with full keyboard navigation and external screen readers support that integrates with existing BBPEs. Our work addresses this need.

\begin{table*}[htbp]
\centering
\caption{Comparison of Accessible Block-Based Programming Approaches}
\label{tab:corrected_comparison}
\setlength{\tabcolsep}{3pt}
\renewcommand{\arraystretch}{0.95}

\begin{tabular}{@{}lllllllp{2.4cm}@{}}
\toprule
\textbf{System} & \textbf{Architecture} & \textbf{Navigation} & \textbf{Editing} &
\textbf{Testing} & \textbf{AT Strategy} & \textbf{Platform} & \textbf{Evaluation} \\
\midrule
Google Text Blockly~\cite{koushik2016accessible} & Blockly Modification & Text-based & N/A & N/A & Screen Reader & Web & Not reported \\
\addlinespace[0.5em]
Ludi et al.~\cite{ludi2017design} & Design Proposal & Full Keyboard & Limited & N/A & Screen Reader & Web & No \\
\addlinespace[0.5em]
Accessible Blockly~\cite{mountapmbeme2022accessible} & Blockly Modification & Partial Keyboard & Mostly & Limited & Screen Reader & Web & User study (n=12)\textsuperscript{1} \\
\addlinespace[0.5em]
Das et al.~\cite{das2021accessible} & Blockly Modification & Partial Keyboard & Limited & N/A & Synth Speech & Web & User study (n=4)\textsuperscript{2} \\
\addlinespace[0.5em]
Tabassum et al.~\cite{tabassum2024accessible} & Blockly Modification & Partial Keyboard & Limited & Limited & Synth Speech & Web & User study (n=18)\textsuperscript{3} \\
\addlinespace[0.5em]
Blocks4All~\cite{milne2018blocks4all} & New System & Touch & Mostly & Yes & iOS VoiceOver & iPad + Robot & User study (n=6)\textsuperscript{4} \\
\addlinespace[0.5em]
OctoStudio~\cite{bulovic2024designing} & New System & Touch & No & Yes & Mobile SRs & Mobile & In progress \\
\addlinespace[0.5em]
Quorum Blocks~\cite{stefik2024accessible} & New Language & Full & Full & Yes & Screen Reader & Cross-platform & User study (n=26)\textsuperscript{5} \\
\midrule
\textbf{EAF (This Work)} & \textbf{Extension} & \textbf{Full Keyboard} & \textbf{Full} &
\textbf{Yes} & \textbf{Screen Reader} & \textbf{Web} & \textbf{Expert eval (n=4)} \\
\bottomrule
\end{tabular}

\vspace{4pt}

\begin{minipage}{\textwidth}
\footnotesize
\raggedright
\textsuperscript{1} One VI student and 17 sighted students.\quad
\textsuperscript{2} 4 BVI participants.\quad
\textsuperscript{3} 12 blind/low-vision programmers.\quad
\textsuperscript{4} 5 children with visual impairments (ages 5--10) and 1 TVI.\quad
\textsuperscript{5} Focus group 1: 5 adults; focus group 2: 21 high school students.
\end{minipage}

\end{table*}

\section{Design Goals}
\label{sec:design}

Based on the identified key limitations of accessibility solutions for BBPEs discussed in Section~\ref{sec:related_work}, we have established five key design goals for the EAF that are outlined below.

\textbf{Structured Navigation.} 
The design should provide structured navigation enabling users to traverse program structures efficiently, aligning with mental models of program structure and block hierarchy. All blocks and elements should be keyboard accessible.

\textbf{Screen Reader Compatibility.} All visual information and interactive states should be conveyed through external screen readers, avoiding built-in synthetic speech that conflicts with users' configured assistive technologies. The design must follow WAI-ARIA~\cite{craig2009accessible} roles, states, and live-region policies to prevent duplicate announcements and ensure focus stays intact.

\textbf{Fully Accessible Editing.} 
Keyboard interactions should provide equivalents for all visual drag-and-drop operations, ensuring mouse-keyboard functional parity.



\textbf{Inclusive Design.} The design should maintain parity between sighted and BVI learners, ensuring that features integrate seamlessly without degrading the experience of any user group.

\textbf{Architectural Modularity.} The framework should seamlessly integrate with existing Blockly-based platforms through a modular architecture, avoiding Blockly library modifications or platform-specific forks. This approach ensures compatibility across Blockly versions and diverse platforms (e.g., Scratch, Code.org).


\section{Extension-Based Accessibility Framework}
\label{sec:architecture}

\subsection{Architecture}
Based on the design goals stated in Section~\ref{sec:design}, we propose an Extension-Based Accessibility Framework (EAF) for Blockly-based programming environments. 
The architecture follows four principles: (1) The framework integrates seamlessly with existing BBPEs without library modifications, ensuring stability. (2) The framework provides WAI-ARIA compliant audio feedback for screen reader compatibility. (3) The framework provides configurable extension points allowing adopting platforms to customize keyboard shortcuts, screen reader verbosity, and color schemes through programmatic API. (4) The framework supports both pointing devices and keyboard interaction, enabling users to choose their preferred input method without compromising functionality.





\begin{figure}[h]
    \centering
    \includegraphics[width=\linewidth]{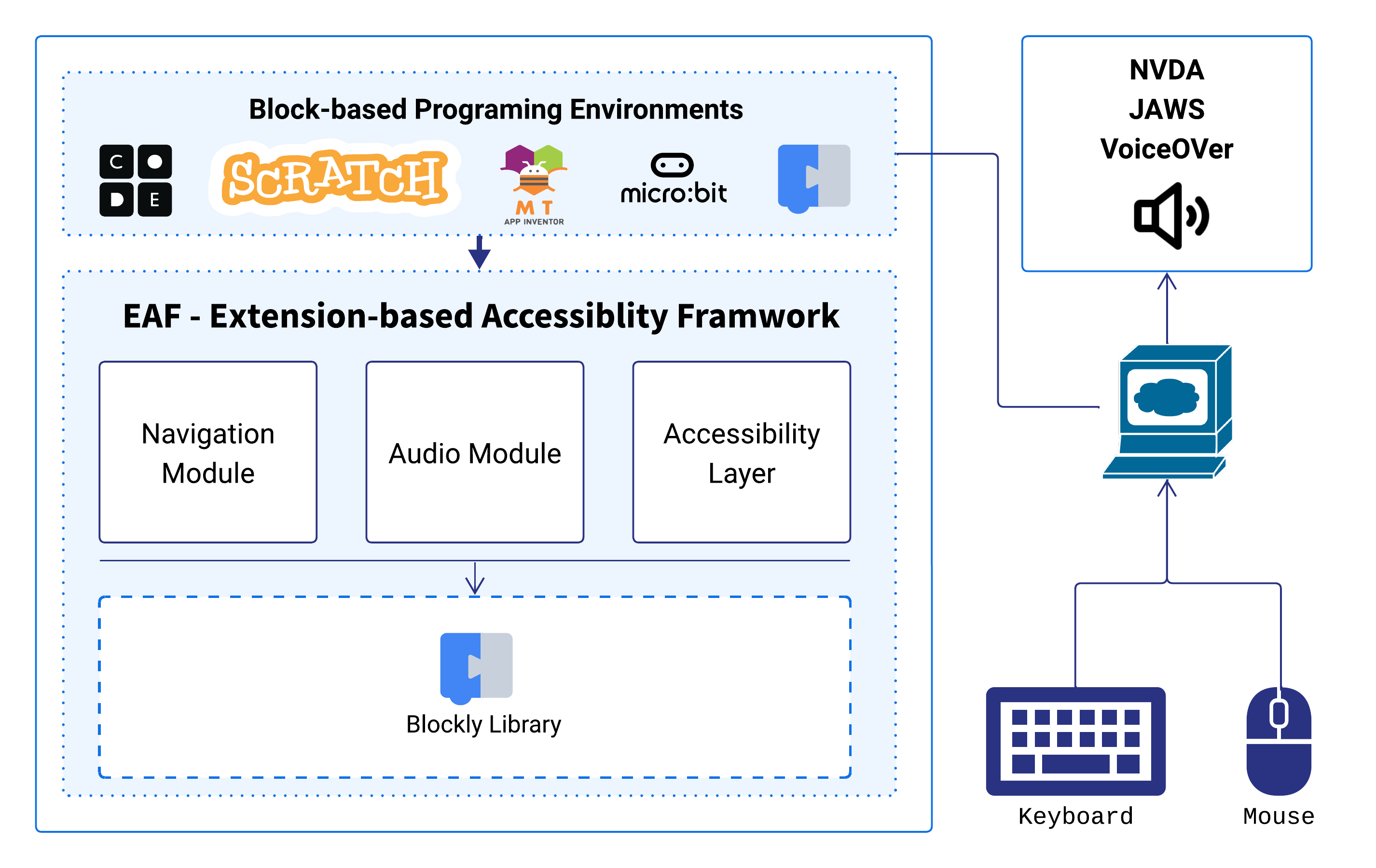}
    \caption{Extension-based Accessibility Framework (EAF) Architecture}
    \label{fig:eaf-architecture}
    \Description{Figure: Extension-based Accessibility Framework (EAF) Architecture}
\end{figure}

The Figure~\ref{fig:eaf-architecture} illustrates the architecture of our framework. The framework is composed of four primary, interconnected components that leverage the browser environment. 

\textbf{Blockly Library}: Blockly library~\cite{google_blockly_library} is used as an association, instead of using it as a dependency of the framework. Our framework shares the same Blockly core semantics and components such as Block, Connection, Input, Field, Toolbox, and Workspace. This approach ensures compatibility with existing block-based platforms that are built on Blockly.

\textbf{Navigation Module}: This module implements an alternative navigational model to support drag-and-drop-like operations via keystrokes and provides comprehensive support for keyboard-only interaction. It also defines standardized keyboard shortcuts for common operations, including block insertion, movement, and traversal. It includes a programmatic API such as custom cursors, markers, and event listeners for extending navigational capabilities.

\textbf{Audio Module}: The audio module generates voice feedback and auditory cues as an alternative output for each keyboard interaction during programming. 
It monitors user interactions via event bus and provides context-based audio feedback. It applies WAI-ARIA roles, states, and properties for seamless integration with VoiceOver, JAWS, and NVDA, providing semantic information about workspace elements.

\textbf{Accessibility Layer}: The framework introduces a suite of accessibility features through its accessibility layer. This layer implements features such as stack labeling, block numbering, and stack jumping functionality to improve accessibility of block-based programming understanding to BVI users to match their sighted peers. These features will help to improve code comprehension and enable faster navigation for BVI users. The accessibility layer provides well-defined extension points, allowing adopting platforms to implement additional accessibility features into Blockly tailored to their specific user needs and domain contexts.

\subsection{Implementation}
The EAF is designed following the JavaScript plugin architectural pattern~\cite{syeed2015pluggable,filippov2023building}, which enables shippable artifacts to host Block-based platforms. The framework is implemented using standard web technologies to ensure broad compatibility and ease of adoption. The core technology stack leverages JavaScript (ES6+) as the programming language, SVG for rendering accessible visual elements within the Blockly workspace, HTML5 for structural markup, and CSS3 for styling and visual feedback. This technology stack aligns with Blockly's native implementation, facilitating seamless integration without introducing additional runtime dependencies. The framework utilizes npm (Node Package Manager)~\cite{npm-package-ref} as the build and distribution system, enabling developers to integrate EAF into projects through standard package management workflows. 

Screen reader support is implemented through comprehensive WAI-ARIA~\cite{craig2009accessible} annotations. The framework dynamically generates and updates \emph{aria-label} attributes based on the current state and context of workspace elements such as blocks, fields, inputs, and connections. ARIA live regions are employed to announce dynamic changes, such as block insertions, deletions, and connection changes, ensuring that screen reader users receive immediate feedback for their actions. The framework has been tested and validated with major screen readers, including VoiceOver (macOS)~\cite{apple_voiceover_mac_2025}, JAWS (Windows)~\cite{freedom_scientific_jaws_docs_2025}, and NVDA (Windows)~\cite{nvaccess_nvda_user_guide_2025_3}.
\section{EAF-based Blockly Design}
\label{sec:accessible_blockly}


\subsection{Code Navigation and Editing Model}
The navigational models proposed in earlier works, shown in Table~\ref{tab:corrected_comparison} are semi-structured or incompletely designed. A structured navigational model is essential to support predictive navigation and better code comprehension, editing, and debugging.

\begin{figure}[h!]
    \centering
    \includegraphics[width=\linewidth]{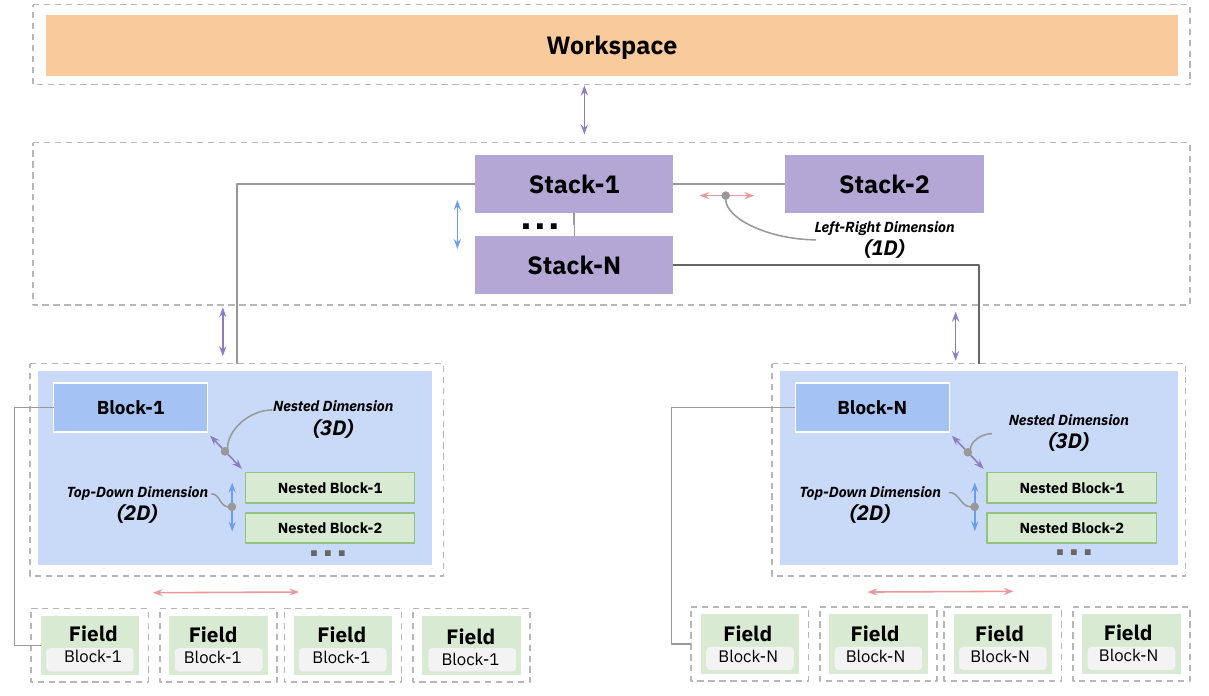}
    \caption{3D Navigation Model Abstraction and Layers}
    \label{fig:5d-abs-layer}
    \Description{Figure: 3-dimensional Navigation Model Abstraction and Layers}
\end{figure}

\textbf{3D Navigational Model.} We have designed a novel Three-Dimensional (3D) navigation model for BVI learners based on programming abstractions such as loops, conditions, variables, etc, and layers such as workspace, stack, blocks, and fields in the Blockly workspace shown in Figure~\ref{fig:5d-abs-layer}. The layer-based traversal and abstractions create contextual boundaries during navigation, which help users orient and maintain context while interacting non-visually with complex code structures~\cite{albusays2017interviews}. This navigation approach ensures less cognitive load while aligning with the cognitive principles of traditional programming~\cite{hassan2023evaluating}.

\begin{figure}[h!]
    \centering
    \includegraphics[width=.8\linewidth, trim=0 10 0 10, clip]{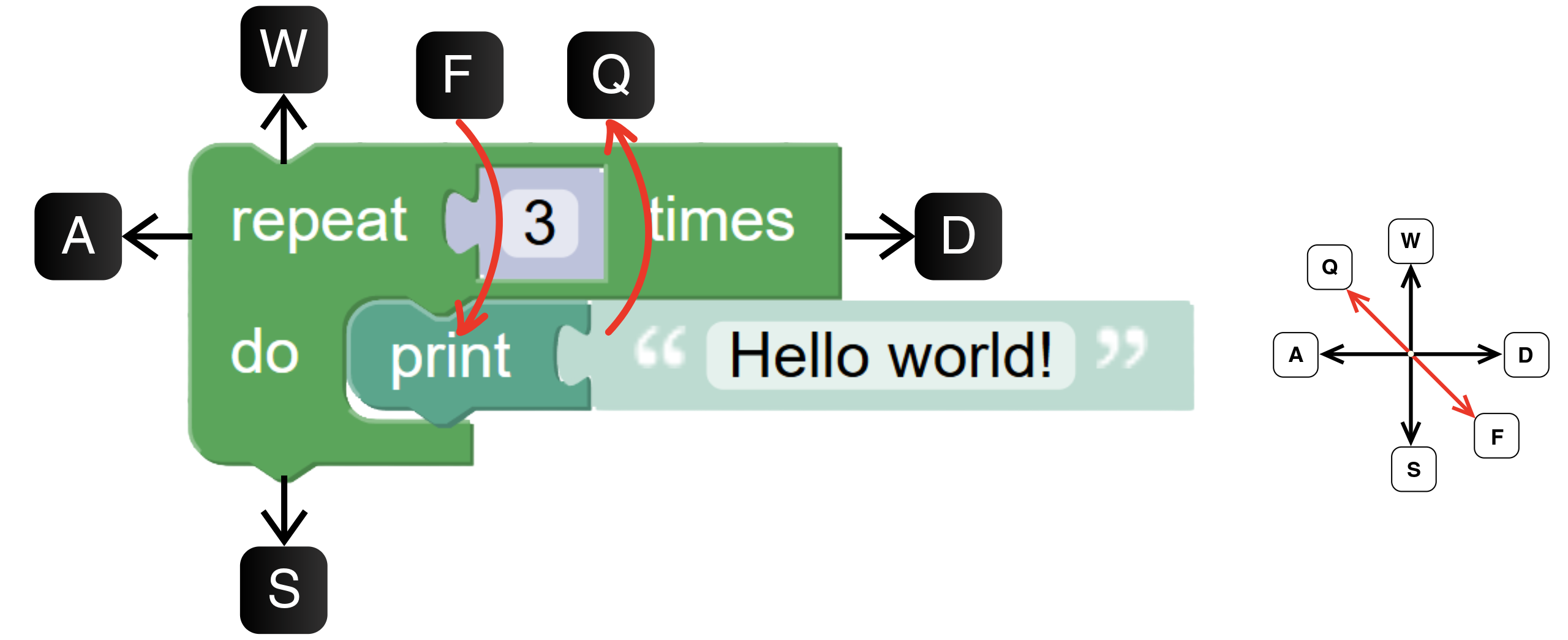}
    \vspace{-0.8em}
    \caption{3D Navigation Model with Keys}
    \label{fig:5d-keys}
    \Description{Figure: 3-dimensional Navigation Model with Keys}
\end{figure}

The 3D model defines three navigation dimensions: [1D] left-right (for horizontal movement within the same hierarchical level); [2D] up-down (for vertical movement within the same hierarchical level); and [3D] in-out (for traversing nesting levels). To implement this model through keyboard interactions, we have adopted the \emph{WASDFQ} key configuration, where \emph{WASD} keys control spatial navigation left, right, up, and down, respectively, and \emph{F} and \emph{Q} keys control nesting in and out, respectively. We selected \emph{WASD} keys over traditional arrow keys because they frequently conflict with default screen reader shortcuts~\cite{ludi2017design} and are consistent with previous attempts on accessible programming solutions~\cite{mountapmbeme2022accessible,okafor2022voice,tabassum2024accessible}. Moreover, the spatial proximity of WASD and FQ keys enables one-handed operation, which may benefit learners with motor impairments by reducing the physical effort for navigation \cite{stefik2024accessible}. Figure~\ref{fig:5d-keys} illustrates the 3D navigation model.


\begin{figure}[htp]
    \centering
    \includegraphics[width=\linewidth]{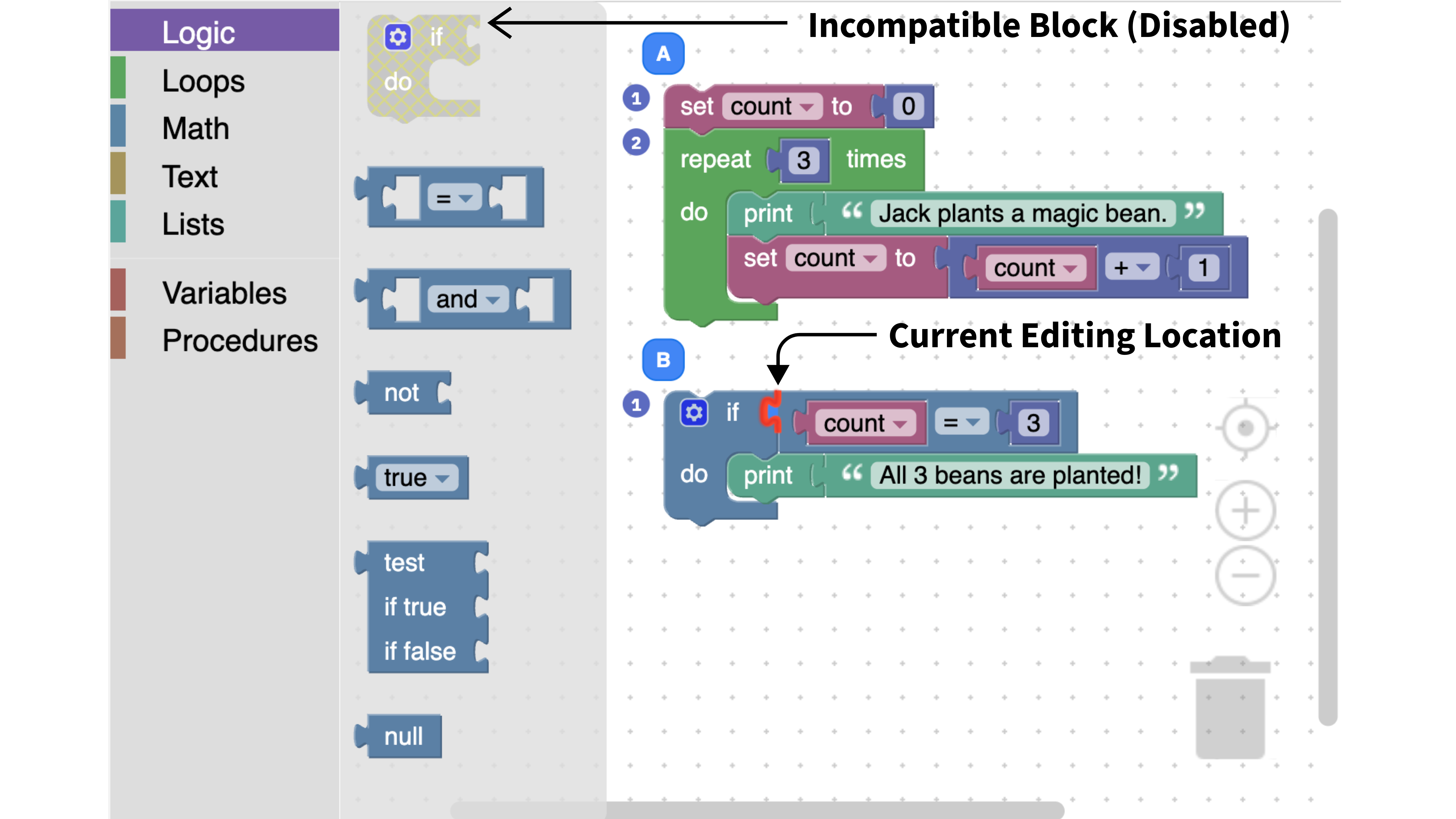}
    \caption{Context-aware Edit Mode}
    \label{fig:edit-mode}
    \Description{Figure: Context-aware Edit Mode}
\end{figure}

\textbf{Editing Mode.} 
Our design uses a mode-based editing approach in which users can switch between editing and navigation modes explicitly. Editing mode activates when a user selects a specific block, at which point modifications or insertions can be performed. This "select first, then edit" mechanism reduces cognitive load by clearly separating navigation and editing tasks~\cite{milne2018blocks4all}, which keeps BVI learners from making unintentional modifications~\cite{mountapmbeme2022accessible}. The editing mode provides keyboard-only alternatives for drag-and-drop operations (attach, detach, cut, copy, paste, delete). In order to reduce distractions, the model uses context-aware editing to show only compatible blocks in the toolbox and hide incompatible ones (Figure~\ref{fig:edit-mode}). Editing mode can be toggled using the \emph{E} key.

\vspace{-15px}

\begin{figure}[h!]
    \centering
    \includegraphics[width=\linewidth]{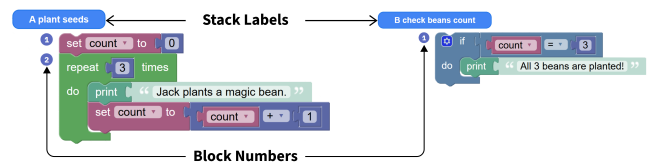}
    \caption{EAF Blockly Workspace with stack labels and block numbers}
    \label{fig:eaf-blockly}
    \Description{Figure: EAF Blockly Workspace with stack labels and block numbers}
\end{figure}

\subsection{Accessible Features}
We have designed a few features to improve the accessibility of block-based programs through our proposed framework. These implementations reside within the accessibility layer of the EAF framework. These features serve as exemplars that developers and researchers of Blockly-based platforms can adapt and extend their systems to improve accessibility according to the needs.

\textbf{Stack Labeling and Block Numbers.} 
The framework introduces stack labeling, assigning unique identifiers to each stack (a group of connected blocks separated from other blocks). By default, stacks are labeled alphabetically (A, B, C, etc.), with customizable labels reflecting semantic purpose (e.g., "main loop"). Block numbers within stacks enable unique block identification. These labels and numbers allow BVI users to reference specific blocks or stacks during navigation. Additionally, stack jumping allows direct navigation to any labeled stack from anywhere in the workspace. For example, if a user is on Stack A and presses \emph{Alt+D}, the cursor jumps from Stack A to Stack D, significantly improving navigation efficiency in complex programs.

\textbf{Navigational Assistant.}
We developed a navigational assistant that provides predictive movement information. When activated with \emph{Shift+H}, the assistant informs users where their cursor will move if they press specific keys. This feature is helpful when users are unsure of their current context or have forgotten shortcuts, helping them learn shortcuts and navigate with confidence.

\textbf{Cursor Locator}
A common frustration for BVI learners is losing track of cursor position after accidentally pressing keys or touching the mouse~\cite{utreras2020accessibility,mountapmbeme2022accessible}. To address this, we have implemented a cursor locator feature. When users press the \emph{C} key, it provides voice feedback of the current position with the current and surrounding blocks' information. 

\textbf{Zoom Controls}
Zoom controls are not currently accessible via keyboard in Blockly~\cite{das2021accessible}, which presents a significant barrier for low-vision users. Our framework implements keyboard interactions for zoom controls: the \emph{+} key zooms in, the \emph{-} key zooms out, and the \emph{0} key resets the view to the default zoom level.




\subsection{Standard Keyboard Shortcuts}
Prior implementations used custom keyboard shortcuts that vary across platforms, creating confusion when users transition between environments and increasing the learning curve. This lack of standardization limits interoperability. Our framework establishes a standardized keyboard shortcut schema serving as a baseline for Blockly-based environments (Table~\ref{tab:shortcuts})

\definecolor{rowgray}{gray}{0.96} 
\newcommand{\key}[1]{\texttt{#1}}

\begin{table}[!htbp]
\caption{Keyboard shortcuts for EAF}
\label{tab:shortcuts}
\scriptsize
\begingroup
\setlength{\tabcolsep}{6pt}
\renewcommand{\arraystretch}{1.35}
\rowcolors*{2}{white}{rowgray}

\centering
\begin{tabular}{@{}l l p{0.55\columnwidth}@{}}
\toprule
\textbf{Scope} & \textbf{Shortcut} & \textbf{Action} \\
\midrule
Navigation   & \textbf{\key{W}}            & $\uparrow$ Move up to the previous block \\
Navigation   & \textbf{\key{A}}            & $\leftarrow$ Move left to the previous block \\
Navigation   & \textbf{\key{S}}            & $\downarrow$ Move down to the next block \\
Navigation   & \textbf{\key{D}}            & $\rightarrow$ Move right to the next block \\
Navigation   & \textbf{\key{F}}            & Move to the first nested block or to the first nested connection \\
Navigation   & \textbf{\key{Q}}            & Move out to the parent block or the outer layer \\
Navigation   & \textbf{\key{Alt+[A--Z]}}   & Jump to a stack labeled with that letter \\
Mode         & \textbf{\key{E}}            & Toggle Edit mode (enter or exit) \\
Workspace    & \textbf{\key{Shift+W}}      & $\uparrow$ Move cursor up \\
Workspace    & \textbf{\key{Shift+S}}      & $\downarrow$ Move cursor down \\
Workspace    & \textbf{\key{Shift+A}}      & $\leftarrow$ Move cursor left \\
Workspace    & \textbf{\key{Shift+D}}      & $\rightarrow$ Move cursor right \\
Toolbox      & \textbf{\key{T}}            & Open toolbox \\
Toolbox      & \textbf{\key{Esc}}          & Close toolbox; focus workspace \\
Announce     & \textbf{\key{C}}            & Speak/announce cursor location \\
Edit ops     & \textbf{\key{Ctrl+X}}       & Cut selected block (Nav) \\
Edit ops     & \textbf{\key{Ctrl+C}}       & Copy \\
Edit ops     & \textbf{\key{Ctrl+V}}       & Paste detached block to current connection (Edit) \\
Edit ops     & \textbf{\key{Delete}}       & Delete selected block \\
Edit ops     & \textbf{\key{Ctrl+/}}       & Add/Hide comment on selected block (Edit) \\
Edit ops     & \textbf{\key{Shift+X}}      & Disconnect at the current cursor location (Edit) \\
Assist       & \textbf{\key{Shift+H}}      & Toggle navigational assistant \\
Assist       & \textbf{\key{Shift+K}}      & Toggle keyboard shortcuts list \\
Assist       & \textbf{\key{Shift+I}}      & Customize stack label \\
Execution    & \textbf{\key{Shift+R}}      & Run the program \\
Execution    & \textbf{\key{Shift+O}}      & Access the output \\
Settings     & \textbf{\key{Ctrl+Shift+K}} & Enable or disable keyboard accessibility \\
\bottomrule
\end{tabular}
\endgroup
\end{table}

To evaluate the proposed accessibility features and demonstrate the framework's integration capabilities, we developed a custom BBPE built on the Blockly library, integrated with the EAF framework as a reference implementation, which is available on GitHub~\cite{eaf-blockly-environment}.
\section{Evaluation}
\label{sec:evaluation}


\subsection{Methodology}
We conducted a two-phase evaluation process combining the technical verification and expert assessment. The methodology involved integration testing of the EAF followed by a semi-structured interview with 4 participants. The goal is to validate system integration, functional stability, and compatibility with screen readers across heterogeneous systems to test and evaluate accessibility features.

\textbf{Integration Testing.} We followed an integration testing~\cite{ipate1997integration} method to evaluate both system-level and feature-level integration of EAF with Blockly-based programming environments. For system-level integration, we employed EAF on two Blockly-based systems: EAF Blockly and an open-source Blockly-based environment to verify the interoperability across different implementations. For feature-level testing, we prepared a comprehensive test suite with 177 test cases targeting all accessibility features with various navigational and editing scenarios. Each test case represents a user action, such as navigating cursors, moving, or inserting blocks. Our testing process covered screen readers such as VoiceOver, JAWS, and NVDA in macOS and Windows operating systems. Then the test cases were assigned among three experienced software developers who participated as testers in the whole testing process. 

\textbf{Semi-structured Interview.} We conducted a task-based evaluation comparing the default keyboard navigation with EAF. The objective was to assess the improvements in the navigation flow, screen-reader feedback, and overall user experience within the BBPEs. Sessions were conducted remotely via Zoom. Participants were provided with supporting materials, including setup documentation, video tutorials, and keyboard shortcuts list to familiarize themselves with the environment before the interview.

Each session lasted 120 minutes and was divided into two segments. The first 30 minutes focused on exploring the Blockly default keyboard navigation. Because this version lacks screen reader support, participants mainly focused on keyboard navigation and orientation. Participants performed three tasks: Navigation (exploring the workspace and locating blocks using keyboard), Create Program (building simple logic sequences to assess structural comprehension), and Edit Program (modifying values, rearranging blocks, and verifying feedback). The remaining 90 minutes were used for EAF with structured 3D navigation, stack labels, block numbers, and editing features. During this phase, participants performed all the tasks using their preferred system with screen reader and keyboard-only interactions and screen readers as stated in Table~\ref{tab:participants}. Feedback for this phase was gathered following completion of all activities.

\subsection{Participants and Demographics}
We recruited four participants (P1-P4) for human evaluation through accessibility research communities. 
All participants provided informed consent, and the study was approved by our institutional review board (IRB).
The group consists of two accessibility experts and two undergraduate computer science students, representing the range of programming experience and familiarity with screen readers. 
The expert participants had extensive familiarity with the accessibility tools and practices, while the undergraduate participants represented novice to intermediate users of block-based programming. All participants were sighted, as the evaluation focuses on technical validation of the EAF framework and accessibility design. Table~\ref{tab:participants} summarizes participants.

\begin{table}[!t]
\centering
\renewcommand{\arraystretch}{1.1}
\setlength{\tabcolsep}{4pt}
\scriptsize

\caption{Summary of Participant demographics\\[-2pt]}
\label{tab:participants}

\begin{tabular}{@{}
l
p{0.22\columnwidth}
p{0.13\columnwidth}
p{0.11\columnwidth}
p{0.18\columnwidth}
p{0.12\columnwidth}
@{}}
\toprule
\textbf{ID} & \textbf{Level} & \textbf{Prog.\ Exp.} & \textbf{SR Fam.} & \textbf{Device / OS} & \textbf{SR Used} \\
\midrule
P1 & Accessibility Expert & Expert        & Moderate & Mac (macOS)     & VoiceOver \\
P2 & Undergrad            & Beginner      & None     & Laptop (Win 11) & NVDA \\
P3 & Undergrad            & Intermediate  & None     & PC (Win 11)     & JAWS \\
P4 & Accessibility Expert & Expert        & Minimal  & Laptop (Win 11) & JAWS \\
\bottomrule
\end{tabular}

\end{table}

Each participant used their personal computers running either a Mac with VoiceOver or a Windows 11 laptop/PC with JAWS or NVDA. All participants received the preparatory documents beforehand to ensure familiarity with the tools and interview process. 

\subsection{Results} 

\subsubsection{Integration Test Result.}
\label{sec:integration-results}

We tested the EAF with two open-source Blockly-based environments. The system-level integration was successful, and all EAF modules were functional. System-level validation confirmed full module compatibility without any functional conflicts. Feature-level testing consisted of n=177 individual test cases (see Table~\ref{tab:test_summary}), covering navigation and movement, mode controls, block editing, stack labeling, accessibility, and general interface functions. Among these 139 tests (78.5 \%) passed successfully, while 38 tests (21.5 \%) failed initially due to minor issues like label misalignment, shortcut conflicts, and missing ARIA attributes. Subsequent debugging verified 13 valid defects, which were corrected during post-testing refinement, raising the overall pass rate to 152 of 177 tests (85.9\%). All core accessibility features, such as 3D navigation, stack jumping, stack labeling, and mode-based editing, remained fully functional across both macOS and Windows environments. Testing with screen readers confirmed reliable keyboard-only support and consistent screen reader compatibility, showing that EAF achieved stable cross-platform integration and satisfied all accessibility performance goals.

\begin{table}[!t]
\centering
\renewcommand{\arraystretch}{1.1}
\setlength{\tabcolsep}{4pt}
\scriptsize

\caption{Categorized feature-level testing result\\[-2pt]}
\label{tab:test_summary}

\begin{tabular}{@{}p{0.40\columnwidth} c c c c @{}}
\toprule
\textbf{Category} & \textbf{Tests} & \textbf{Pass} & \textbf{Fail} & \textbf{Bugs Fixed} \\
\midrule
Navigation \& Movement                & 38  & 35 & 3  & 3 \\
Mode Controls \& Interface Management & 28  & 19 & 9  & 4 \\
Block Editing \& Operations           & 28  & 28 & 0  & 0 \\
Stack Labeling, Search \& Numbering   & 54  & 30 & 24 & 6 \\
Accessibility \& Screen Reader        & 10  & 8  & 2  & 0 \\
Zoom \& View Controls                 & 9   & 9  & 0  & 0 \\
Help \& System Features               & 10  & 10 & 0  & 0 \\
\midrule
\textbf{Total Tests} & \textbf{177} & \textbf{139 (78.5\%)} & \textbf{38 (21.5\%)} & \textbf{13 (85.9\%)} \\
\bottomrule
\end{tabular}

\end{table}


\subsubsection{Participants' Feedback} 
Participant feedback and behavioral observations were analyzed thematically across four dimensions, namely navigation, creating/editing a program, screen-reader interaction, and recommendations for enhancement.

\textbf{Navigation Experience} Across all sessions, participants reported clearer spatial orientation and less confusion when using EAF compared to the default keyboard navigation. During the default keyboard navigation, P1 struggled to move within the nested blocks, describing it as "\textit{trial and error, sometimes impossible to know what is going on,"} while P4 mentioned it as \textit{"very confusing and feels inconsistent."} and said it lacks feedback when actions fail. In the EAF, P1 successfully moved between the nested print block and a repeat block and observed that "\textit{the navigation was a lot better with F and Q it was clearer and easy".} P4, while traversing multiple stacks, highlighted that the framework \textit{"is more tied to program scope,"} and \textit{"build a mental model that you can move in and out of easily."} Among the undergraduates, P2 mentioned that the new shortcut layout was \textit{"easy to remember,"} and P3 described the stack-jumping feature was \textit{"very helpful, much easier than the other one."}

\textbf{Editing and Program interaction} During editing tasks, participants shared their experience with Blockly's default keyboard navigation and EAF's keyboard navigation.
In the default keyboard navigation editing task, P1 struggled with feedback uncertainty while attempting to cut and paste blocks, stating that \textit{"the system status is completely unknown for me. It's not telling me what I can or cannot do."} In the EAF, P1 completed edits seamlessly, reflecting that \textit{"it was good, not hard."} Similarly, P2 faced challenges with editing tasks in the default keyboard navigation but was able to add and edit blocks in the EAF environment, commented \textit{“by exploring more, I can do better."} P3 used Edit mode to modify variable blocks, describing the process as \textit{"pretty easy"}. P4 performed multiple actions, such as rearranging, cutting, and reattaching blocks within the edit mode, and summarized the workflow as \textit{"a lot easier to navigate"}. 

\textbf{Screen Reader Feedback} Participants reported that EAF integration with screen readers like JAWS, NVDA, and VoiceOver produced detailed and reliable auditory feedback. P4 observed that \textit{"the screen reader modifies what's being presented based on what you're actually doing. I really like that. It is very specific as to what block or stack you're connected to"}. P2 described the output as \textit{"helpful for navigation"}. With VoiceOver, P1 noted that  \textit{"the text was too much, the reading was a bit slow, I wish the voice was faster"} and P3 found JAWS \textit{"too fast and jumbled"}. 

\textbf{Participant Recommendations} Participants suggested shorter speech output, clearer indicators for mode changes, and more simplified onboarding materials. During the editing phase, P1 noted that \textit{"each time, I didn't know whether I was in edit mode; it wasn't completely clear even after using it for a few times,"} further adding to that \textit{"students without a screen reader cannot understand it clearly"}. P2 stated that, \textit{"If I practice again, I can remember"}. The experts emphasized feedback improvement, with P1 noting that \textit{"the text was too much, the reading was bit slow, I wish the voice was faster"} also added \textit{"the system status is completely unknown, it is not telling me what I can do or cannot do"} and P4 suggested \textit{"a quick corrective noise or something to tell you that you are doing something wrong"}.


\section{Discussion}
\label{sec:discussion}


\subsection{Findings}
Our evaluation indicates that the EAF successfully addresses the fundamental challenges of making Google Blockly accessible to BVI learners without modifying the library code. The integration testing confirmed the structural stability across different Blockly-based environments, and participants' feedback offered profound insight for improving keyboard navigation with screen readers.

\textbf{Navigation Model. }
The positive feedback regarding EAF's keyboard-only navigation indicated that 3D navigation model aligns with the conceptual structure of the programming rather than solely its visual spatial arrangement. P4's observation \textit{("build a mental model that you can move in and out of easily")} suggests that the hierarchical navigation model successfully mapped to participants' understanding of program structure (loops, conditions, nesting). This differs from default navigation, which participants characterized as \textit{"trial and error, sometimes impossible to know what is going on."} The predictability of EAF's directional model seems to make it easier for participants to understand the spatial orientation and reduce cognitive load. This addresses a major limitation found in previous research on accessible block-based - programming.~\cite{mountapmbeme2022accessible,tabassum2024accessible}. The fact that undergraduate participants found the shortcuts "easy to remember" while expert participants found them "tied to program scope" shows that the navigation model works for users with different levels of expertise.

\textbf{Screen Reader Integration. }
The experiences of participants with VoiceOver, JAWS, and NVDA demonstrate that EAF's WAI-ARIA implementation effectively facilitates compatibility with external screen readers. P4 stated that the feedback was "\textit{very specific as to what block or stack you're connected to,}" and P2 stated it was "\textit{helpful for navigation.}" This demonstrates voice-based audio feedback are an effective way to inform users about the program structure and navigation status. However, the evaluation also showed trade-offs that were specific to screen readers. VoiceOver gave P1 "\textit{too much}" information, and P1 stated that the reading was "a bit slow." P3 noted that JAWS was "\textit{too fast and jumbled.}" These differences show that the primary issue is not that the two systems are fundamentally incompatible, but that they need to be adjusted to fit the user's speech rate and verbosity. Future improvements could prioritize user-configurable verbosity settings over adaptations tailored for specific screen readers. This method avoids the differences between internal synthetic speech and external screen readers documented in previous research~\cite{das2021accessible}.

\textbf{Extension-Based Architecture. }
Our architectural approach is proven by the fact that EAF works well with two Blockly-based environments without having to modify the core library. Previous accessibility solutions either required forking Blockly's codebase~\cite{das2021accessible} or creating new systems~\cite{tabassum2024accessible}. EAF's extension-based design adds accessibility while being compatible with new Blockly releases. Existing BBPEs such as Scratch, MakeCode, and Code.org can potentially adopt EAF without restructuring their current implementations, lowering the barrier to widespread accessibility adoption. The framework's ability to run on browsers makes it cross-platform compatible and accessible from anywhere. 


\textbf{Comparison to Prior Approaches. }
EAF's design directly addresses key limitations identified in related work. Where Accessible Blockly~\cite{das2021accessible} required core modifications and lacked screen reader compatibility, EAF maintains architectural separation while supporting external screen readers. Where Blocks4All~\cite{milne2018blocks4all} focused on simplified tasks, EAF supports full Blockly functionality including nested structures. Where default keyboard navigation lacked structure, EAF's hierarchical approach provides predictability. The combination of modularity, screen reader compatibility, and structured navigation represents a synthesis of best practices from prior accessibility research applied to the Blockly ecosystem.

The success of EAF's extension architecture demonstrates that accessible BBPEs need not sacrifice either functionality or modularity. The rapid task completion and positive feedback suggest that keyboard-based block programming, when properly structured, can be viable for BVI learners despite the inherently visual nature of traditional block-based environments. The findings presented here are subject to the limitations discussed in Section~\ref{sec:limitation}.

\subsection{Takeaways}
We have derived the following takeaways for researchers and developers working with BVI users to make BBPEs inclusive. 

\textbf{Accessibility Extensions.} Extension-based approaches like EAF can offer a practical pathway to improving accessibility in existing BBPEs. By providing a shippable, modular solution, it can enable accessibility without requiring platforms to implement accessibility support from scratch. This approach can lower the barrier to the adoption of accessible BBPEs in classrooms.

\textbf{Standard Keyboard Shortcuts.} Our design establishes a standard set of keyboard shortcuts for BBPEs. These standards can help BVI users to develop transferable skills and switch between platforms without needing to relearn shortcuts. Broad adoption of proposed standards could foster a more inclusive and interoperable ecosystem of BBPEs.

\textbf{Accessible UI Elements.} Sighted users intuitively perceive visual cues in UI components such as blocks and stacks to comprehend code, which are inaccessible to BVI users. Enhancements to Blockly UI components, such as stack labels and block numbers, can help BVI learners better understand program structure and comprehend code. We recommend that researchers and developers explore designing block-based UI elements with accessibility in mind that compensate for visual cues for BVI users.
\section{Limitations}
\label{sec:limitation}
Our study demonstrates the viability of EAF, but we acknowledge several limitations that should be addressed in future studies.

\textbf{Evaluation Scope.} We evaluated the framework involving experts. The accessibility experts offered key insights into screen reader compatibility, interaction patterns and overall accessibility design, we lack direct feedback from BVI participants. Also, all participants were adults, while most BBPEs targets K-12 learners. Testing with BVI students in real classroom settings would provide critical understandings into the EAF's effectiveness and usability.

\textbf{Platform Validation.} We designed EAF to work with any Blockly-based platform. We tested the integration using open-source Blockly distributions. However, EAF compatibility with public BBPEs such as MakeCode or Code.org remains unknown, as the system is proprietary.

\textbf{Assistive Technology Coverage.} We successfully validated EAF with three major screen readers, but our testing did not include refreshable braille displays, which some BVI programmers prefer. We tested keyboard-based zoom controls, however, how it renders with screen magnification software for low-vision users needs further evaluation.

Despite these limitations, our results indicate that EAF architecture can make BBPEs accessible for BVI users while maintaining platform compatibility and integrity for widespread adoption, a strategy that has been largely overlooked in previous studies.


\section{Conclusion and Future Work}
\label{sec:conclusion}
This study presents the Extension-Based Accessibility Framework (EAF) for Google Blockly. The frameworks successfully integrate with Blockly-based environments without requiring modifications to Blockly library. The results indicate that it makes Blockly-based BBPEs fully accessible to BVI learners through keyboard-only navigation and support for external screen readers. Our design introduces a 3D navigation model with accessible block UI elements and mode-based context-aware program editing. Our evaluation and feedback from accessibility experts and computer science students validate that these efforts help to improve code comprehension, understanding program structure, accessibility of code navigation, and editing. 

In our future work, we will conduct comprehensive user studies engaging BVI learners to assess the EAF's effectiveness in block-based programming learning in classroom settings. Additionally, our current work focuses on the EAF architecture and accessible code navigation and editing models. In future work, we will explore accessible debugging methods and techniques for making graphical outputs, such as animations and interactive visualizations, accessible to BVI users.


\begin{acks}
We thank Google for funding this work through the Blockly Accessibility Fund initiative.
\end{acks}

\bibliographystyle{ACM-Reference-Format}
\bibliography{ref}

\end{document}